# Assessing transportation accessibility equity via open data

Amirhesam Badeanlou*[1], Andrea Araldo[2], Marco Diana[3]

[1] Student, DIATI, Politecnico di Torino, Italy
[2] Associate Professor, Institut Politechnique de Paris, Télécom SudParis, France
[3] Associate Professor, DIATI, Politecnico di Torino, Italy

**SHORT SUMMARY**

We propose a methodology to assess transportation accessibility inequity in metropolitan areas. The methodology is based on the classic analysis tools of Lorenz curves and Gini indices, but the novelty resides in the fact that it can be easily applied in an automated way to several cities around the World, with no need for customized data treatment. Indeed, our equity metrics can be computed solely relying on open data, publicly available in standardized form. We showcase our method and study transportation equity of four cities, comparing our findings with a recently proposed approach.

**Keywords:** Transportation accessibility, Transportation equity, Open Data.

## 1. INTRODUCTION

The use of private cars by the suburban population is among the main causes of pollution (Grelier, 2018 ) and reducing their auto-dependency is the first objective toward sustainability of urban transportation. To pursue the aforementioned goal, mass transit is of paramount importance: the suburban populations will get rid of their current auto-dependency only if they are provided with a competitive mass transit offer, offering good Quality of Service (QoS), so that travelers would find it more convenient than their private cars.

However, although irreplaceable (Basu, Araldo et al., 2018), mass transit is known to be inefficient in the outskirts. In such areas the transportation demand is too low to justify high transit frequency and high transit stop density, which would result in an unfeasible operational cost / passenger. As a consequence, the suburban population generally suffers from waiting and walking-to-station times much higher than in city centers. This inequity (Calabrò, Araldo et al., 2021) is structural in classic mass transit and forces suburban travelers to use private cars (Welch and Mishra, 2013) .

It is thus important to quantify such inequity. An approach that has been widely used is to compute *accessibility* measures. In broad terms, the accessibility of a certain location measures how well it is connected to the rest of the surrounding urban area. Several metrics have been proposed to quantify accessibility (Welch and Mishra, 2013). Transportation equity can be assessed by studying the geographical distribution of the chosen accessibility metric: if there is a big difference between accessibility in the city center and in the suburbs, this is an indication of high inequity.

Most work on accessibility equity assessment requires rich data about the area under study, e.g., households and employment in each zone. As a consequence, in order to study a certain city, one would need to contact the respective authority, hoping to obtain datasets and perform customized data processing on each dataset, due to the lack of a standard format. This process may take months for just one city or might even be infeasible. For this reason, many work on accessibility equity just focus on one or two scenarios.

We propose in this paper a computation methodology to evaluate accessibility equity, only relying on open data, available in standardized form. This allows to compute accessibility equity metrics of multiple cities in an automated way, without the need of time-consuming case-by-case data processing. This allows also to easily compare accessibility equity of different cities.

Our methodology consists in computing *accessibility scores* via third party open-source code (Biazzo, Monechi and Loreto, 2019) and then automatically compute Lorenz curves and gini indices to assess equity of distribution of accessibility. Our code is available as open source (https://github.com/andreaaraldo/public-transport-analysis).

We showcase our methodology comparing accessibility equity in Madrid, Paris, Boston and Sydney. The results we obtain mainly corroborate findings on the same cities from (Biazzo, Monechi and Loreto, 2019). However, some metrics contradict each other, which shows that finding the "right" accessibility equity measures is still open to debate and requires further effort from the scientific community.

## 2. METHODOLOGY

To assess accessibility equity, we first compute accessibility of *hexagons*, and then we plot Lorenz curves, based on such hexagons.

*Accessibility score*
In particular, as in (Biazzo, Monechi and Loreto, 2019), we partition the area under study in hexagons $\lambda \in \Lambda$, each of 0.5Km per side. We compute the **velocity score** $v(\lambda)$ of each hexagon $\lambda$, i.e., the average speed at which it is possible to move starting from the center of the hexagon and going toward a random direction, by using public transportation. To compute the **sociality score** $s(\lambda)$ we need to also have the information about the population density within each hexagon. The sociality score of a certain hexagon $\lambda$ measures how many individuals one can potentially reach, moving away from the center of the hexagon. The exact formulas of $v(\lambda)$ and $s(\lambda)$ are in (Biazzo, Monechi and Loreto, 2019).
Observe that the aforementioned scores would vary with time of day, together with the transit line frequencies. We here use the values averaged over a daily period from 6AM to 8PM.

As an example Figure 1 shows the velocity and sociality scores of Boston. More than the absolute value of the scores, in order to assess equity, we need to focus on the geographical distribution of accessibility. Note that, as expected, both velocity and sociality scores are much better in the city center than the suburbs. Accessibility score maps for the cities considered in this paper are reported in the Appendix. They show the same trend as Boston.

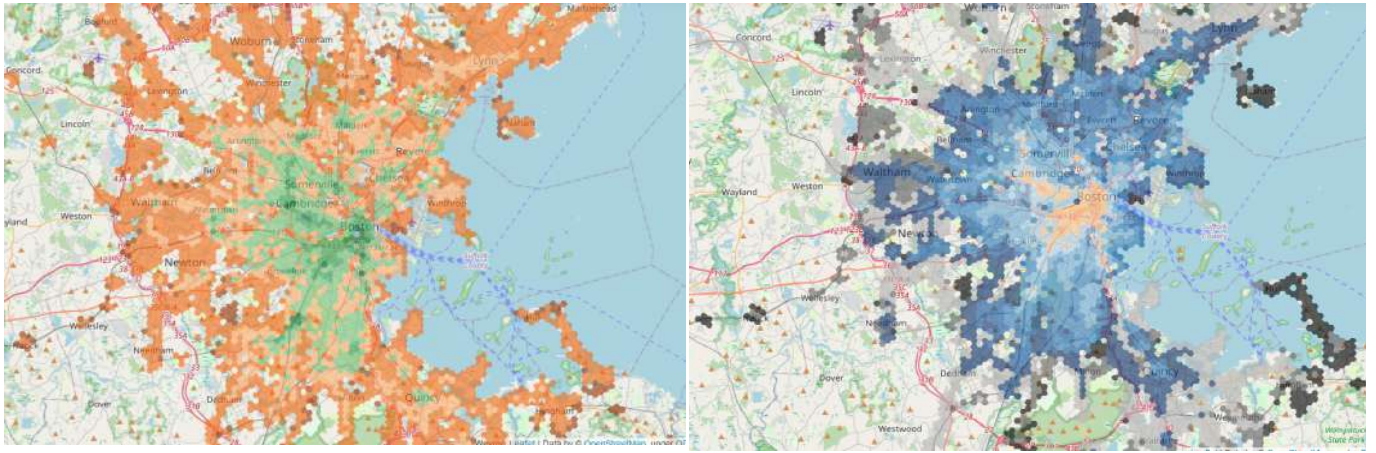

(a) Velocity score  (b) Sociality score

**Figure 1: Accessibility scores for Boston.**

In what follows, we will denote with $a(\lambda)$ the accessibility score of hexagon $\lambda$, which might be either $v(\lambda)$ or $s(\lambda)$, depending on the context.

*Lorenz curve: general description*

In order to assess equity, we resort to Lorenz curves, which allows us to visualize and quantify the fairness with which a certain *resource* is distributed among *stakeholders*. Usually, the considered resource is income and the stakeholders' are individuals. Fig. 1 is an example of Lorenz curve.

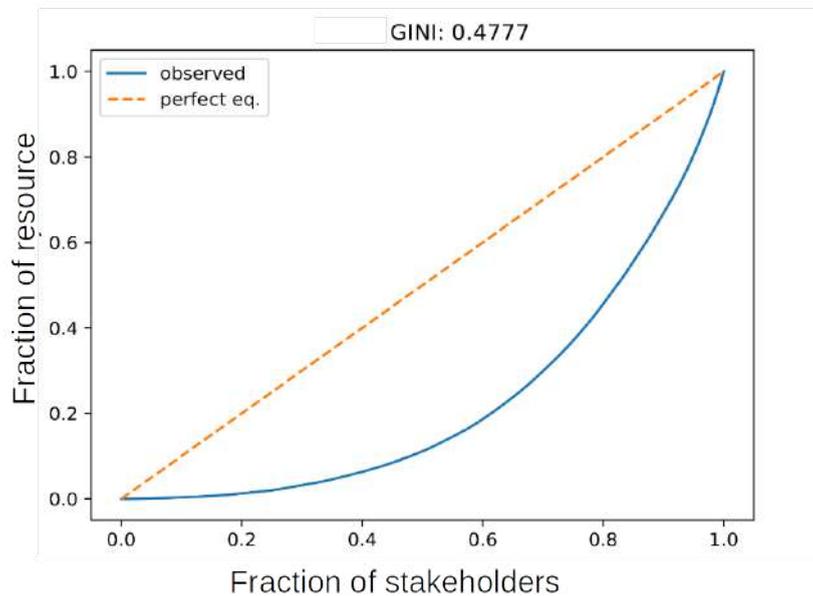

**Figure 2: Example of Lorenz curve**

In the x-axis, we order the stakeholders from the worst (the one that takes the smallest part of resource) to the best. For the sake of normalization, we represent centiles of the stakeholders, instead of directly the stakeholders, so that the x-axis ends in 1. In the y-axis we report the cumulative amount of resources, normalized so that the y-axis ends in 1. If we take the blue Lorenz curve of Figure 2 (the exact interpretation is for the moment not important), we can say that 60% of the stakeholders get less than 20% of the resource. Obviously, in the perfect equity case, we would like that $x$% of population gets $x$% of the resource, $\forall x \in [0, 1]$. This case is represented by the straight orange line.

The smaller the distance between the Lorenz curve and the perfect-equity curve, the more the equity of the distribution of the resource. The distance between the curves is computed via the Gini index, which is obtained by (i) computing the area between the Lorenz curve and the perfect-equity curve and then (ii) dividing it by 0.5 (which is the area below the perfect equity curve). Gini index goes from *0* to *1*, where *0* means that there is perfect equity, while *1* denotes maximal inequity.

*Lorenz curve for accessibility equity*
Lorenz curve has been used for assessing transportation accessibility equity (Welch and Mishra, 2013). In that case, the resource is some measure of transportation quality, while stakeholders might be individuals. Differently from previous work, our method only relies on open data, available in standardized form.

Let us consider the accessibility scores $a(\lambda), \forall \lambda \in \Lambda$ (it can be either velocity or sociality score, i.e., $v(\lambda)$ or $s(\lambda)$). We construct two types of Lorenz curves, where the resource is the accessibility.

The first is the **hexagon-based Lorenz curve**. We consider each hexagon as a stakeholder. We order the hexagons from the worst to the best, so that $a(\lambda_i) \leq a(\lambda_{i+1})$. We put such hexagons $\lambda_1, \cdots, \lambda_{|\Lambda|}$ in the x-axis. For each hexagon $\lambda$, the corresponding value of the Lorenz curve is

$$L_a^{hex}(\lambda_i) = \frac{1}{K} \cdot \sum_{j=1}^{i} a(\lambda_j).$$

Constant *K* is a normalization factor, so that the Lorenz curve goes from *0* to *1*, i.e., $K = \sum_{j=1}^{|\Lambda|} a(\lambda_j)$. We also normalize the x-axis, so that it goes from *0* to *1*.

We also compute the **population-based Lorenz curve**. This time, we consider an individual as a stakeholder. Let us consider an individual $p$ living in hexagon $\lambda$ (we indicate this with $p \in \lambda$). We assume that all the individuals living in any hexagon $\lambda \in \Lambda$ enjoy the accessibility of that hexagon, i.e., $a(p) = a(\lambda), \forall p \in \lambda$. Let us denote with *P* the set of all individuals. We order the individuals $p_1, \cdots, p_{|P|}$, such that $a(p_i) \leq a(p_{i+1})$. Observe that, if we browse individuals in such an order, we will first encounter all individuals in the worst hexagon (the one with the smallest accessibility), then the second worst and so on and so forth. We put the individuals $p_1, \cdots, p_{|P|}$ in the x-axis. For each individual $p_i$, the Lorenz curve is

$$L_a^{ind}(p_i) = \frac{1}{K'} \cdot \sum_{j=1}^{i} a(p_j).$$

Similarly as before, constant $K'$ is a normalization factor, so that the Lorenz curve goes from $0$ to $1$, i.e., $K' = \sum_{j=1}^{|P|} a(p_j)$. We also normalize the x-axis, so that it goes from $0$ to $1$.

Since the accessibility metric can be either velocity or sociality score, we compute 4 types of Lorenz curves in total, as in Table 1.

**Table 1: Types of Lorenz curves**

| Notation | Description | Stakeholders | Accessibility score |
|---|---|---|---|
| $L_v^{hex}(\lambda_i)$ | Hexagon-based Lorenz curve of the velocity score | Hexagons | Velocity |
| $L_s^{hex}(\lambda_i)$ | Hexagon-based Lorenz curve of the sociality score | Hexagons | Sociality |
| $L_v^{ind}(\lambda_i)$ | Individual-based Lorenz curve of the velocity score | Individuals | Velocity |
| $L_s^{ind}(\lambda_i)$ | Individual-based Lorenz curve of the sociality score | Individuals | Sociality |

We denote the Gini indexes computed on the respective Lorenz curves as $G_v^{hex}, G_s^{hex}, G_v^{ind}, G_s^{ind}$.

## 3. RESULTS AND DISCUSSION

*Dataset and code*
We solely used open data listed in Table 2, which are encoded with the same format. We need two kinds of information: transit schedules, to compute journeys within transit, which allows us to understand how fast we can move from a hexagon. With this information, we can compute hexagon-based Lorenz curves. In order to compute the sociality score or the population-based Lorenz curves, we also need to know the population density within each hexagon.

As for the transit schedules, we exploited the General Transit Feed Specification (GTFS), a data standard adopted by all the main transportation authorities around the world to publish their transit schedules.

**Table 2: Open data used**

| Description | Source |
|---|---|
| GTFS data (transit schedules) | transitfeed.com |
| Population density of European cities | EUROSTAT (https://ec.europa.eu/eurostat/web/gisco/geodata/reference-data/grids) |
| Population density of non-European cities | Gridded Population of the World, Version 4 (GPWv4): Population Count. Palisades, NY: NASA Socioeconomic Data and Applications Center (SEDAC); 2016. (doi:10.7927/H4X63JVC). |

*Analysis procedure*

We first compute the accessibility scores using CityChrone software (Biazzo, Monechi and Loreto, 2019) and we visualize their geographical distribution in the map (Figures A1 and A2). We then trace the Lorenz curves and compute Gini indices (Figures A3-A6). We compare Madrid, Paris, Boston and Sydney, which are among the analyzed cities from (Biazzo, Monechi and Loreto, 2019). They quantify the inequity by two metrics:
- The ratio between the average accessibility of the best 1% hexagons and the overall average accessibility or
- The ratio between the average accessibility of the best 1% population and the overall average accessibility

We believe that the aforementioned metrics of inequity are not reliable. Indeed, choosing the 1% best scores seems arbitrary. Indeed, if another percentage is used (say 5%), inequity considerations might be completely different. We therefore choose to quantify inequity as the Gini index calculated on Lorenz curves. Gini index is more general than the metrics of (Biazzo, Monechi and Loreto, 2019), as it does not require to arbitrarily choose any parameter.

The Gini indexes we calculated are reported in Table 3. Recall that the lower the Gini index, the better the equity. To allow for an easier comparison between cities, we normalize each column via min-max scaling and report the results in Table 4. For the sake of easier interpretation, we plot Table 4 in Figure 3.

**Table 3: Gini indexes**

|        | $G_v^{hex}$ | $G_s^{hex}$ | $G_v^{ind}$ | $G_s^{ind}$ |
|--------|-------------|-------------|-------------|-------------|
| Paris  | 0.3868      | 0.4777      | 0.3809      | 0.4259      |
| Madrid | 0.3762      | 0.4436      | 0.3584      | 0.3803      |
| Boston | 0.3721      | 0.4306      | 0.3713      | 0.4166      |
| Sydney | 0.3757      | 0.5208      | 0.371       | 0.4232      |

**Table 4: Normalized gini indexes**

|        | $G_v^{hex}$ | $G_s^{hex}$ | $G_v^{ind}$ | $G_s^{ind}$ |
|--------|-------------|-------------|-------------|-------------|
| Paris  | 0.62        | 0.11        | 0.47        | 0.32        |
| Madrid | -0.10       | -0.27       | -0.53       | -0.68       |
| Boston | -0.38       | -0.42       | 0.04        | 0.11        |
| Sydney | -0.14       | 0.58        | 0.03        | 0.26        |

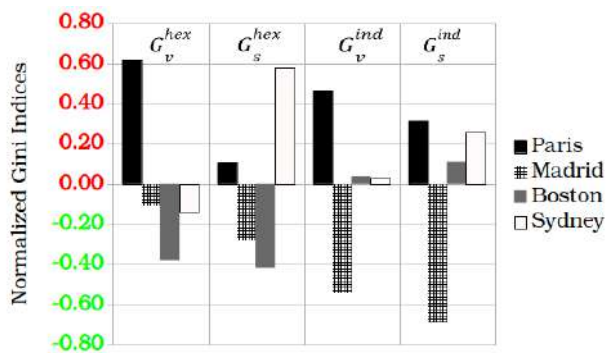

**Figure 3: Normalized Gini indices (the larger, the higher inequity)**

We first notice that Paris suffers the most from high inequity, with most of the 4 Gini used. On the contrary, Madrid enjoys the best equity. This is confirmed by visually comparing the geographical distribution of the accessibility scores from the two cities in Figure 4. It is evident that the accessibility gap between the center and suburbs is way larger in Paris, while in Madrid accessibility is more evenly distributed.

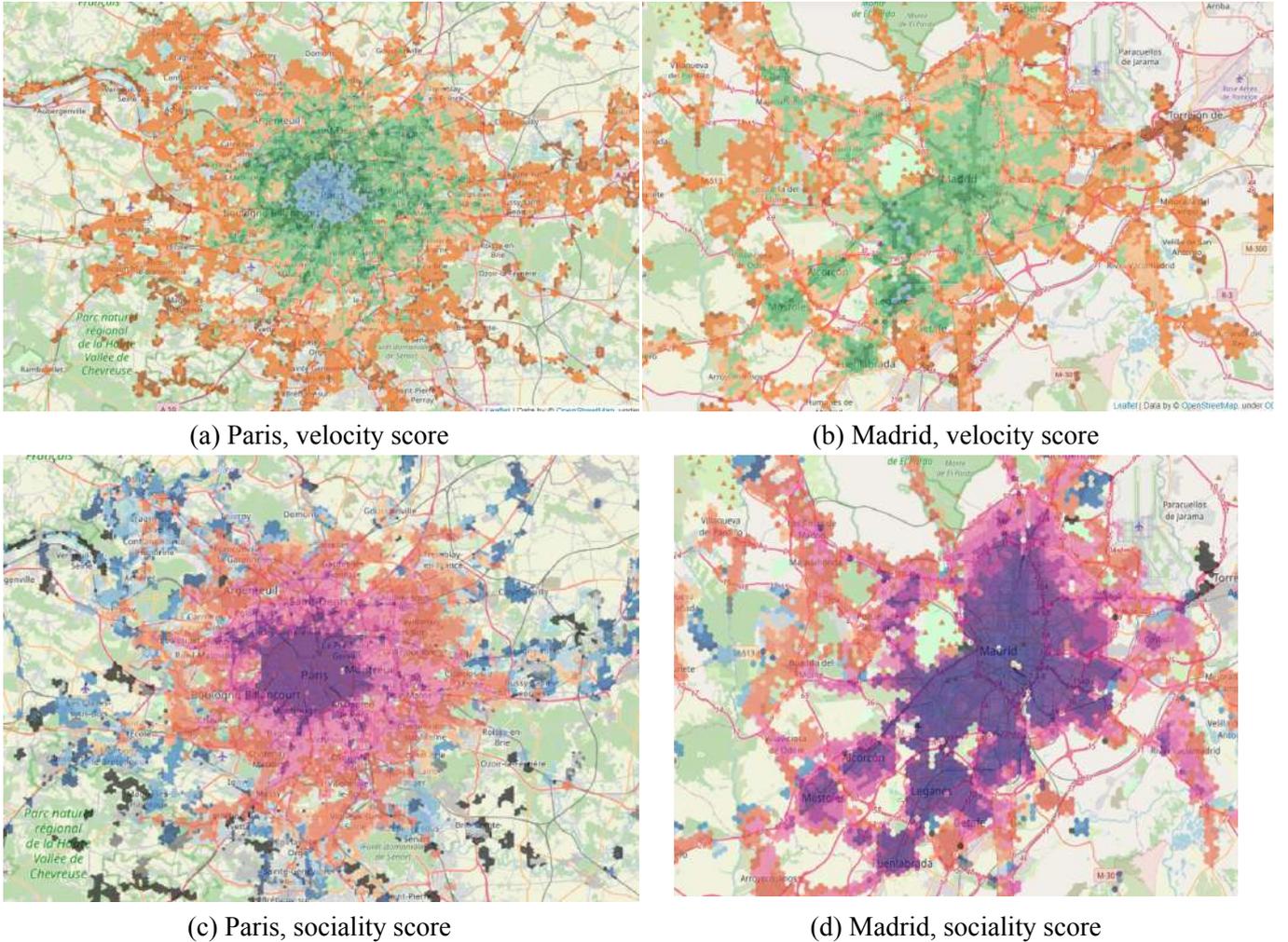

(a) Paris, velocity score  (b) Madrid, velocity score

(c) Paris, sociality score  (d) Madrid, sociality score

**Figure 4: Comparison between the worst city (Paris) and the best (Madrid) in terms of accessibility inequity.**

We can order cities from the worst (high inequity) to the best (high equity), based on the four metrics defined in (Biazzo, Monechi and Loreto, 2019) and based on our four Gini-index based metrics. We do this in Table 5. Note that general trends are confirmed by all metrics, but there are differences. If we focus on the hexagon-based inequity of velocity scores (first two rows of Table 5), based on our metric, Madrid is better than Sydney, in contrast to (Biazzo, Monechi and Loreto, 2019) that claim the opposite. Visual inspection (Figure A1) shows that our claim is correct. In terms of sociality scores (Figure A2), instead, our Gini-index based metrics do not seem to be more accurate than (Biazzo, Monechi and Loreto, 2019).

Overall, one reassuring finding emerges: all metrics manage to capture the very evident differences in accessibility equity, from a city to another. This encourages the possibility to automate equity analysis across cities, with no need of visual inspection, only based on open data.

**Table 5: Comparison of the ranking of cities from (Biazzo, Monechi and Loreto, 2019) and the ranking based on our computation (Figure 3). The latter are in *italic*.**
**Each ranking orders the cities from the one that suffers the highest inequity, to the one that enjoys the highest equity.**
**Green background indicates that the two rankings correspond.**

| Metric | Worst city | 2nd worst | 2nd best | Best |
|---|---|---|---|---|
| Velocity score (top 1% hexagons) | Paris | Madrid | Sydney | Boston |
| $G_v^{hex}$ | *Paris* | *Sydney* | *Madrid* | *Boston* |
| Sociality score (top 1% hexagons) | Paris | Boston | Sydney | Madrid |
| $G_s^{hex}$ | *Sydney* | *Paris* | *Madrid* | *Boston* |
| Velocity score (top 1% individuals) | Paris | Boston | Sydney | Madrid |
| $G_v^{ind}$ | *Paris* | *Boston* | *Sydney* | *Madrid* |
| Sociality score (top 1% individuals) | Boston | Sydney | Paris | Madrid |
| $G_s^{ind}$ | *Paris* | *Sydney* | *Boston* | *Madrid* |

## 4. CONCLUSIONS

We have proposed a methodology to assess the equity on the distribution of transportation quality, which can be fully automated and can be easily performed for several cities across the World. This is guaranteed by the fact the computation only relies on open data in standardized form. We performed a comparative analysis of four cities and we found that our results confirm previous work findings, in some cases better capturing the accessibility inequity differences between cities.

We warn the reader on the fact that results on equity merit a more in-depth analysis: indeed, the gap between the best accessibility hexagon and the worst may depend on the size of the area under study, which is an aspect we did not consider. Moreover, in our future work, in order to assess the realism of the simple metrics we presented here, we will compare them with some other metrics more difficult to compute (requiring more (non-open) data) but also more accurate, for one or two cities for which we may be able to obtain detailed information.

**APPENDIX:**

*Accessibility scores*

Here we report the accessibility score maps that we obtain via CityChrone open source tool (Biazzo, Monechi and Loreto, 2019). The overall trends correspond to the ones reported in the original paper of CityChrone.

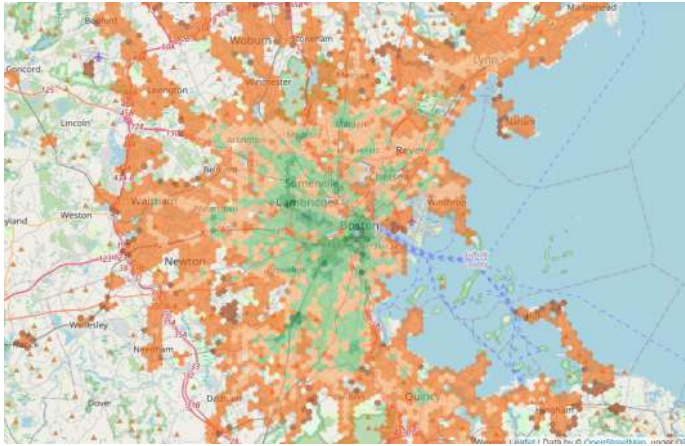
(a) Boston

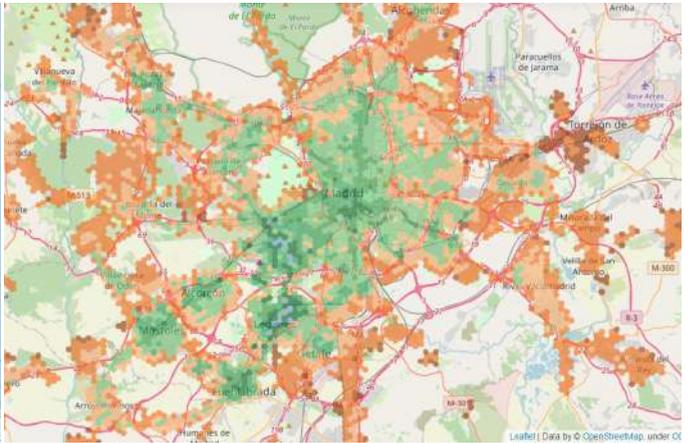
(b) Madrid

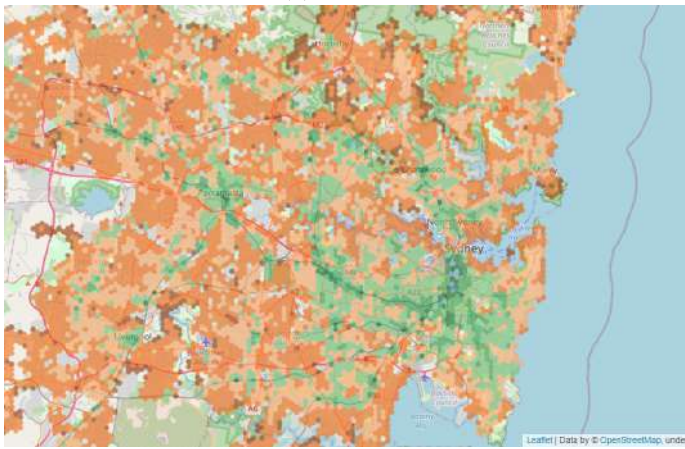
(c) Sydney

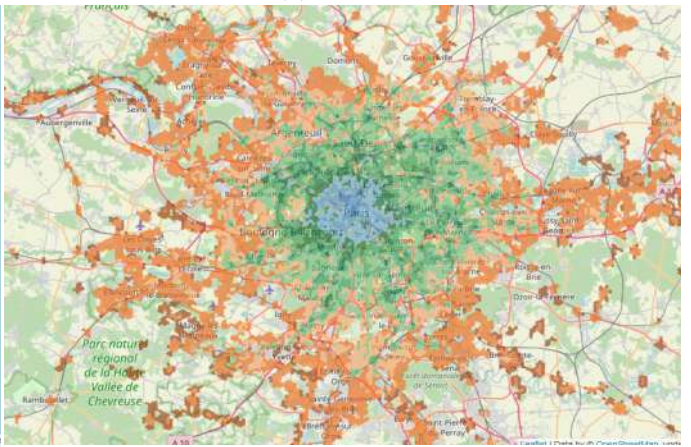
(d) Paris

**Figure A1: Velocity scores**

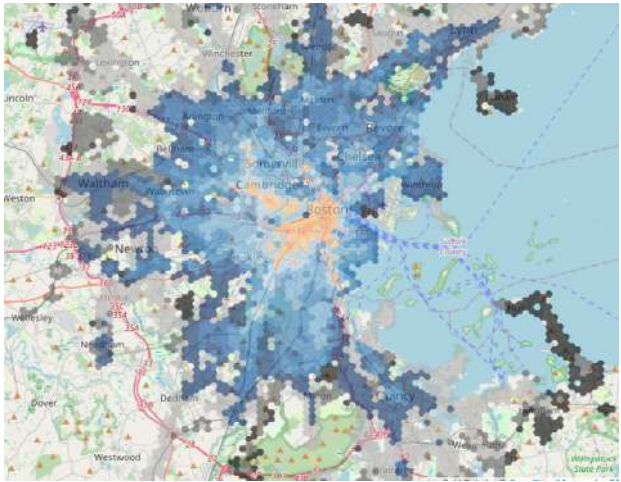
(a) Boston

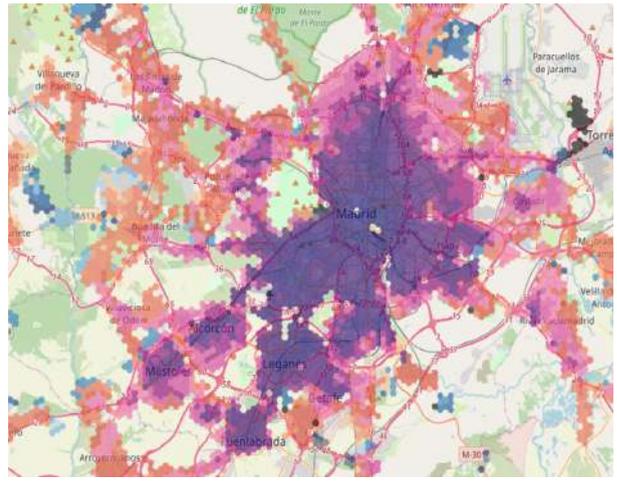
(b) Madrid

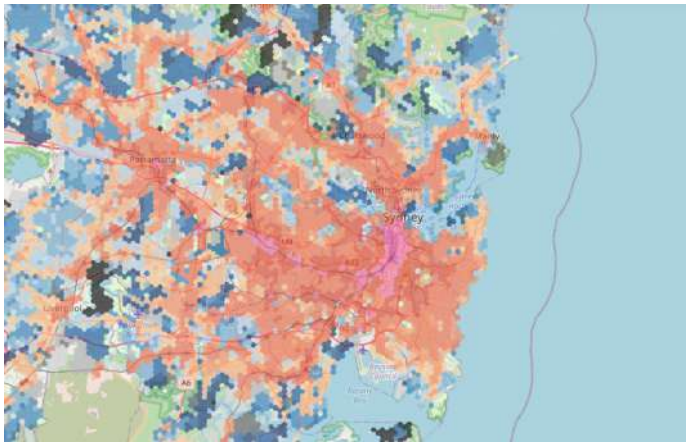
(c) Sydney

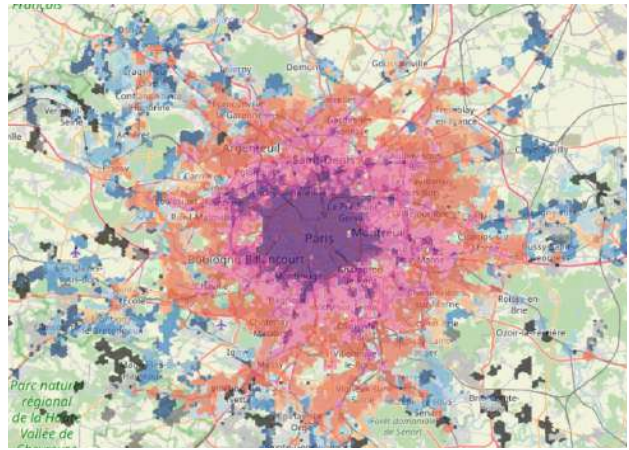
(d) Paris

**Figure A2: Sociality scores**

*Lorenz curves*

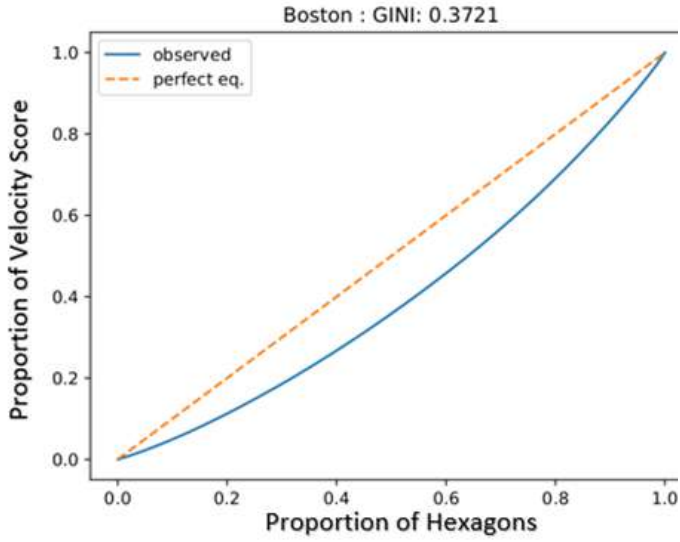
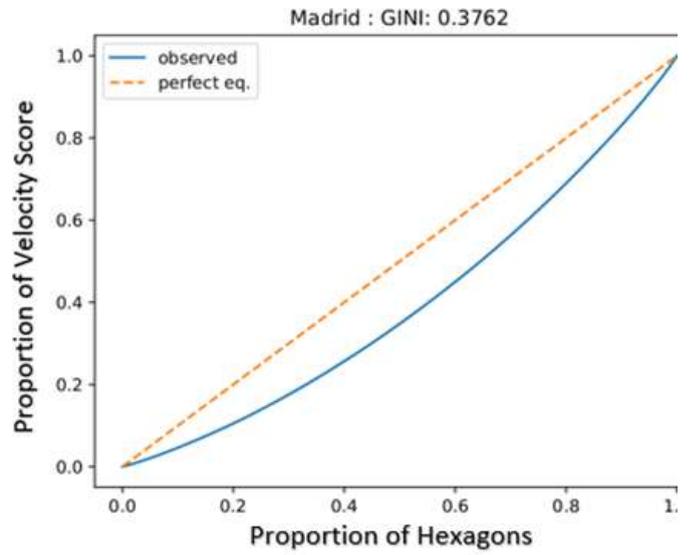

(a) (b)

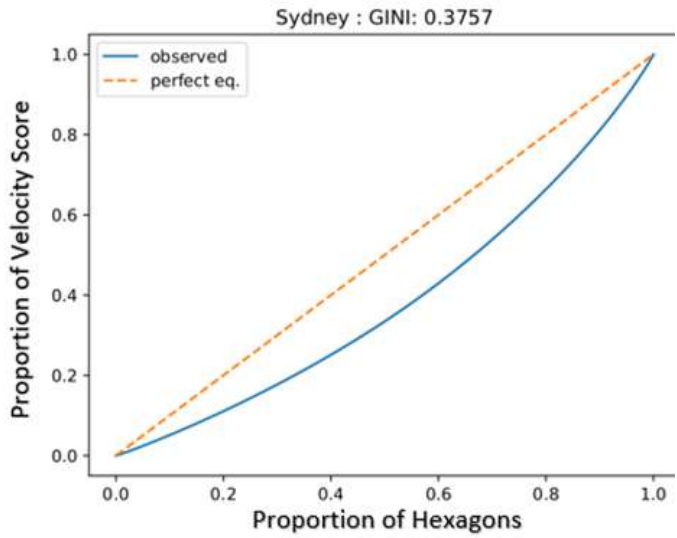
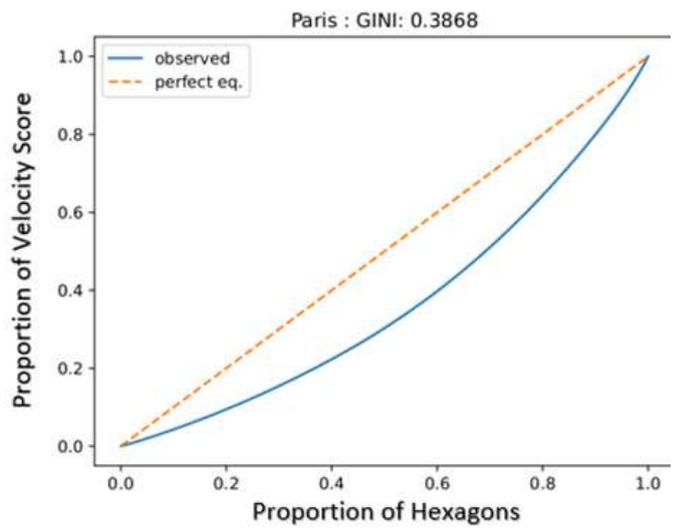

(c) (d)

**Figure A3: Hexagon-based Lorenz curves of the velocity score $L_v^{hex}(\lambda_i)$.**

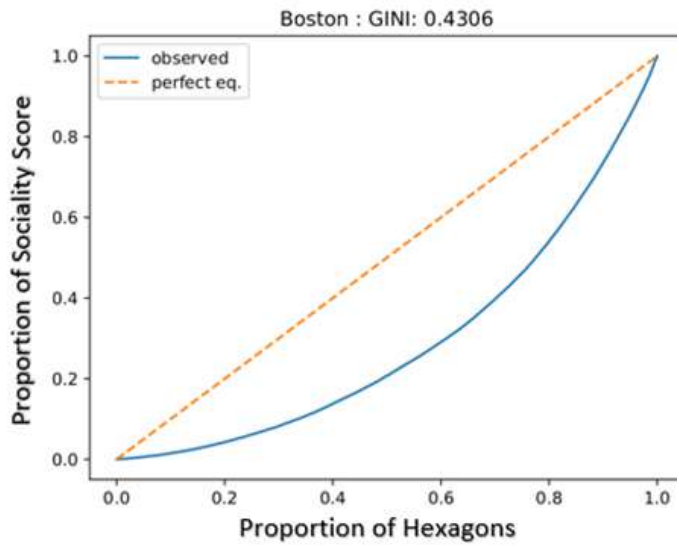
(a)

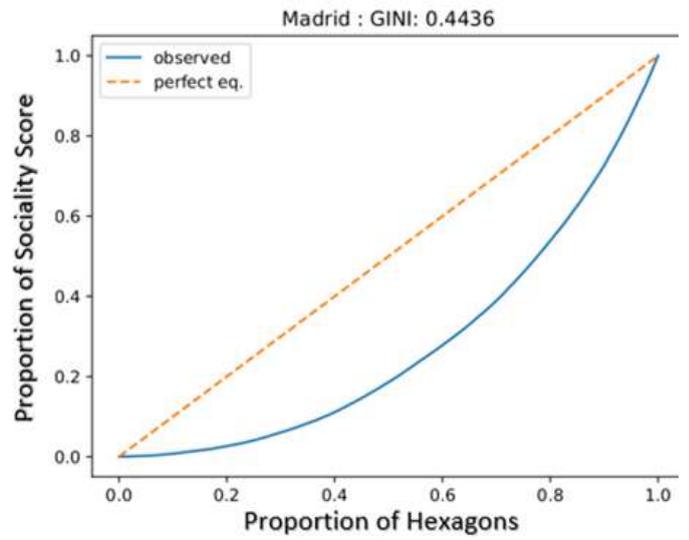
(b)

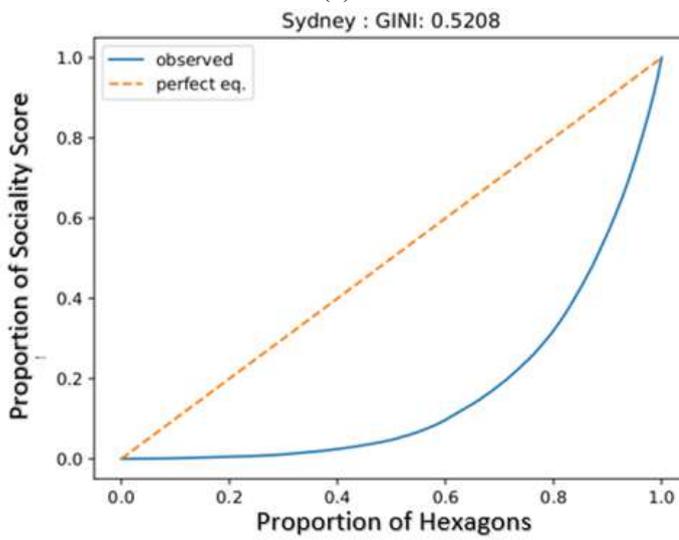
(c)

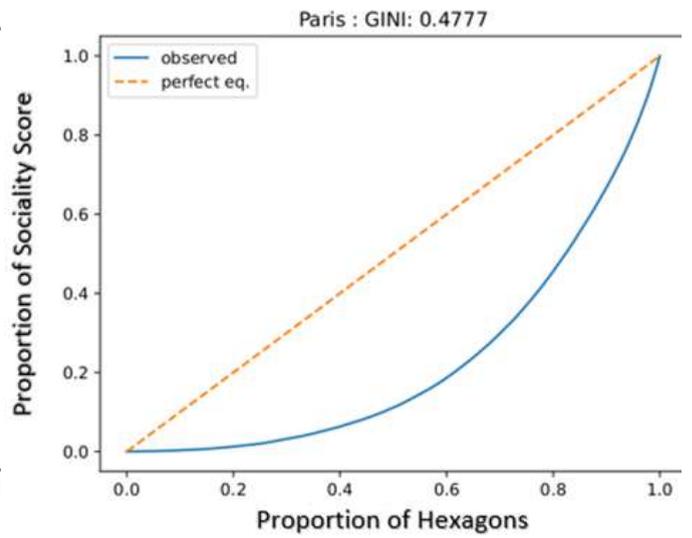
(d)

**Figure A4: Hexagon-based Lorenz curves of the sociality score $L_s^{hex}(\lambda_i)$.**

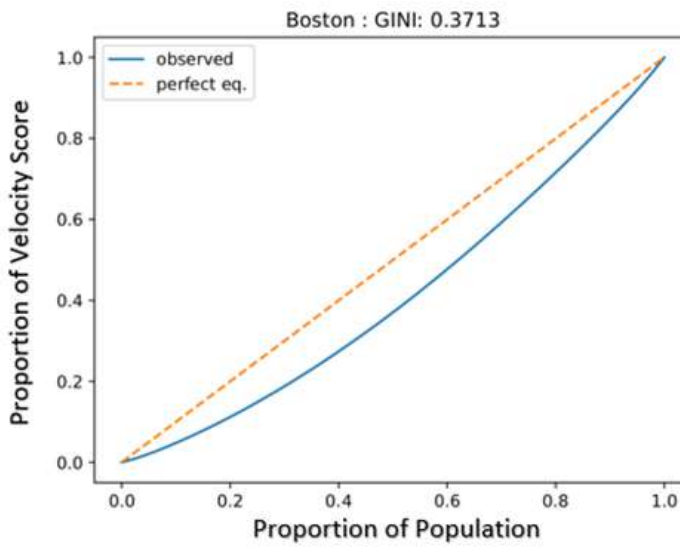
(a)
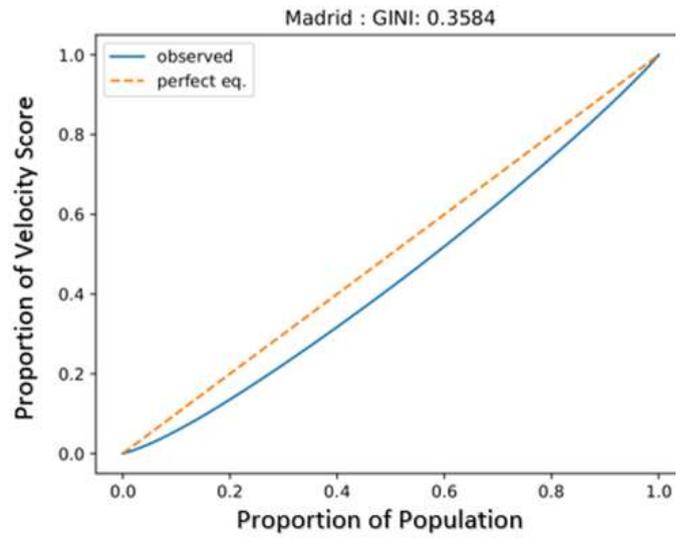
(b)
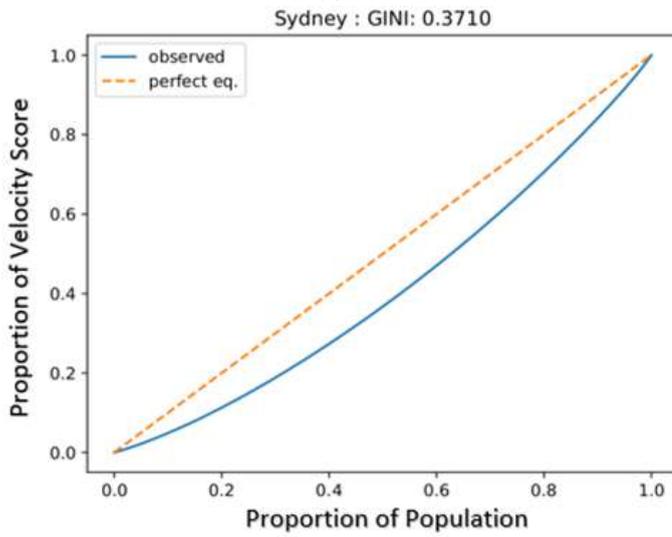
(c)
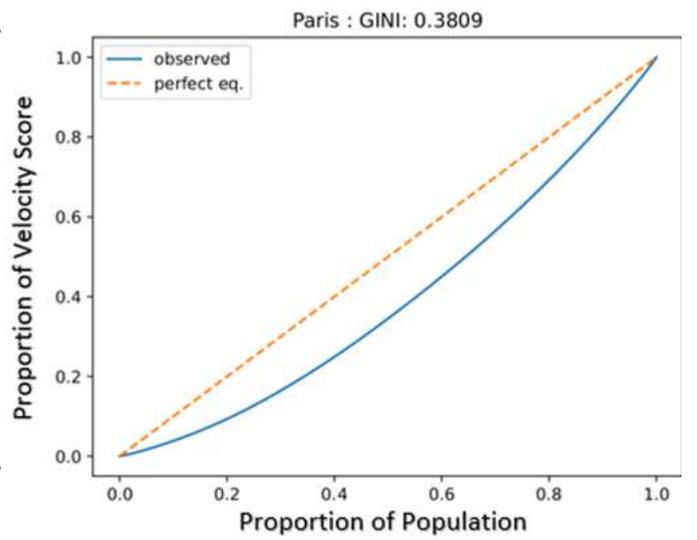
(d)

**Figure A5: Individual-based Lorenz curves of the velocity score $L_v^{ind}(\lambda_i)$.**

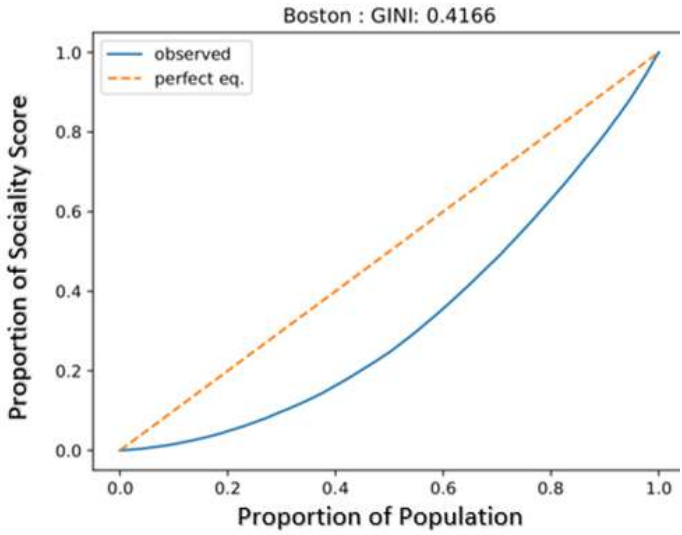

(a)

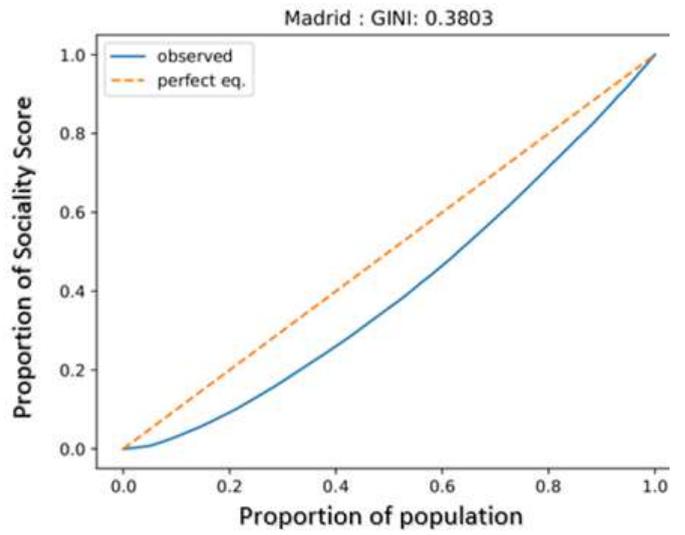

(b)

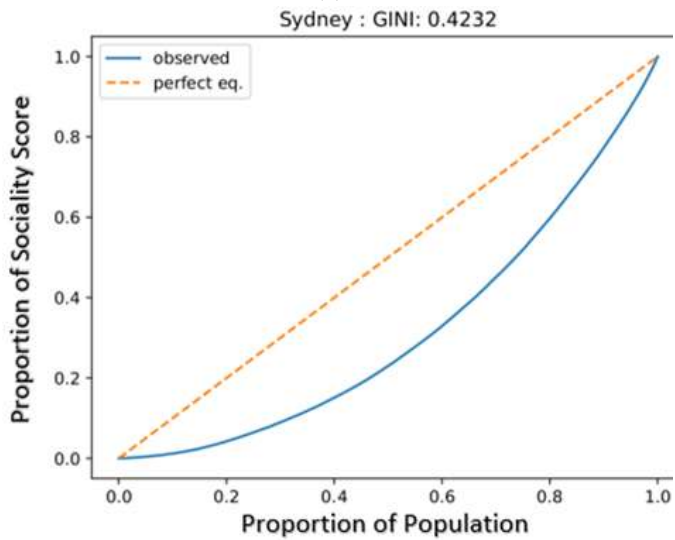

(c)

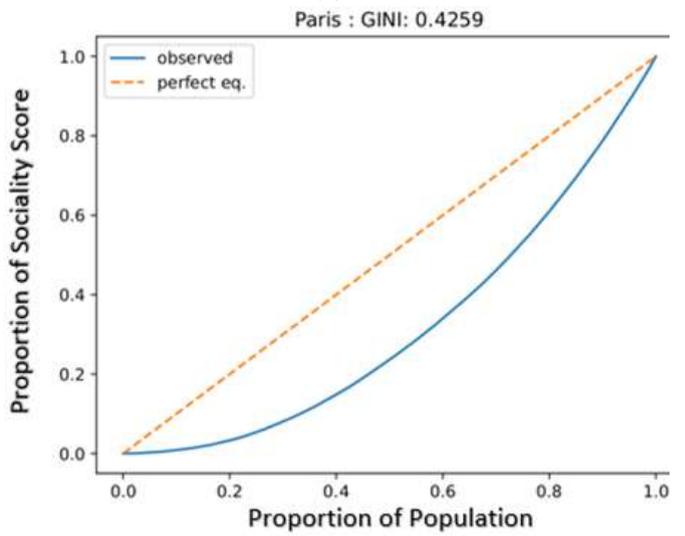

(d)

**Figure A6: Individual-based Lorenz curves of the sociality score $L_s^{ind}(\lambda_i)$.**